\documentclass[twocolumn]{aastex631}

\usepackage{amsmath}
\usepackage{subfigure}
\usepackage{url}
\usepackage{hyperref}
\usepackage{xspace}
\newcommand{\cii}{[C\,{\scriptsize II}]\xspace}
\newcommand{\ci}{[C\,{\scriptsize I}]\xspace}
\newcommand{\hi}{H\,{\scriptsize I}\xspace}
\def\yt{\mbox{\rm{\textsc{yt}}}\xspace}
\def\msun{\mbox{\rm{M}$_{\odot}$}\xspace}
\def\lsun{\mbox{\rm{L}$_{\odot}$}\xspace}
\def\cloudy{\mbox{\rm{C\textsc{loudy}}}\xspace}
\def\sigame{\mbox{\rm{S\textsc{\'{i}game}}}\xspace}
\def\simba{\mbox{\rm{\textsc{simba}}}\xspace}

\def\caesar{\mbox{\rm{\textsc{caesar}}}\xspace}
\shorttitle{The \cii-to-\hi relation at $z \simeq 6$}
\shortauthors{David Vizgan et al.}
\graphicspath{{./}{figures/}}

\begin{document}

\title{Investigating the \cii-to-\hi conversion factor and the \hi\ gas budget of galaxies at $z\approx 6$ with hydrodynamical simulations}

\author[0000-0001-7610-5544]{David Vizgan}
\affiliation{The Cosmic Dawn Center}
\affiliation{National Space Institute, DTU Space,  Technical University of Denmark, 2800 Kgs.~Lyngby, Denmark}
\affiliation{Department of Astronomy, University of Illinois at Urbana-Champaign, 1002 West Green St., Urbana, IL 61801, USA}

\author{Kasper E. Heintz}
\affiliation{The Cosmic Dawn Center}
\affiliation{Niels Bohr Institute, University of Copenhagen, Lyngbyvej 2, DK-2100 Copenhagen, Denmark}

\author{Thomas R. Greve}
\affiliation{The Cosmic Dawn Center}
\affiliation{National Space Institute, DTU Space,  Technical University of Denmark, 2800 Kgs.~Lyngby, Denmark}
\affiliation{Department of Physics and Astronomy, University College London, Gower Place, WC1E 6BT London, UK}

\author{Desika Narayanan}
\affiliation{The Cosmic Dawn Center}
\affiliation{Department of Astronomy, University of Florida, Gainesville, Florida 32611}

\author{Romeel Dav\'e}
\affiliation{Institute for Astronomy, University of Edinburgh, Edinburgh EH9 3HJ, U.K.}
\affiliation{University of the Western Cape, Bellville, Cape Town 7535, South Africa}

\author{Karen P. Olsen}
\affiliation{Department of Astronomy and Steward Observatory, University of Arizona, Tucson, AZ 85721, USA}

\author{Gerg\"o Popping}
\affiliation{European Southern Observatory, 85478 Garching bei M\"unchen, Germany}

\author{Darach Watson}
\affiliation{The Cosmic Dawn Center}
\affiliation{Niels Bohr Institute, University of Copenhagen, Lyngbyvej 2, DK-2100 Copenhagen, Denmark}

\begin{abstract}

One of the most fundamental, baryonic matter components of galaxies is the neutral atomic hydrogen (\hi). At low redshifts, this component can be traced directly through the 21-cm transition, but to infer \hi\ gas content of the most distant galaxies, a viable tracer is needed. We here investigate the fidelity of the fine structure transition of the ($^2P_{3/2} - ^2P_{1/3}$) transition of singly-ionized carbon \cii{} at $158\,\mu$m as a proxy for \hi\ in a set simulated galaxies at $z\approx 6$, following the work by \cite{heintz2021}. We select 11,125 star-forming galaxies from the  simba{} simulations, with far-infrared line emissions post-processed and modeled within the \sigame{} framework. We find a strong connection between \cii{} and \hi, with the relation between this \cii-to-\hi\ relation ($\beta_{\rm [CII]}$) being anti-correlated with the gas-phase metallicity of the simulated galaxies. We further use these simulations to make predictions for the total baryonic matter content of galaxies at $z\approx 6$, and specifically the \hi\ gas mass fraction. We find mean values of $M_{\rm HI}/M_\star = 1.4$, and $M_{\rm HI}/M_{\rm bar,tot} = 0.45$. These results provide strong evidence for \hi\ being the dominant baryonic matter component by mass in galaxies at $z\approx 6$.

\end{abstract}


\section{Introduction}

In the process of galaxy formation, the infall of neutral atomic hydrogen (\hi) is a necessary, fundamental ingredient in the formation of stars. Reservoirs of \hi gas condense into molecular hydrogen (H$_{2}$), which serves as fuel for star formation. As such, \hi is a crucial component for the assembly of galaxies, and thus governs the first epoch of galaxy formation. At higher redshifts (z $>$ 3) \hi can also reveal large-scale structure of the universe by tracing the precursors to superclusters of galaxies \citep[e.g.][]{scott1990}.

The main tracer of \hi is the $\lambda =$ 21 cm line, which arises from transitions between the hyperfine structure levels in the ground state of the hydrogen atom. There are many benefits, in theory, of utilizing the 21 cm line -- for instance, it does not suffer from dust extinction and attenuation like other lines, and kinematic information about the emitting gas can be derived from its Doppler shift \citep[][]{walter2008}. Even though \hi has been observed within our own galaxy \citep[e.g.][]{kalberla2005} and for many galaxies locally \citep[e.g.][]{thuan1981, swaters2002, walter2008}, few detections of the 21-cm line emission have been made beyond $z \approx 0.3$ \citep[][]{bera2019}, mainly due to the weakness of the transition and the limited sensitivity \citep[][]{carilli2013}. The current most distant estimate for a single galaxy has reached $z = 0.376$ \citep[][]{fernandez2016}, with other recent works examining the average 21-cm signal from a stack of a few thousand galaxies out to $z\approx 1$ \citep[][]{chowdhury2020, chowdhury2021}. Although the evolution of the \hi budget across redshift has been modeled via theoretical predictions \citep[][]{madau2014}, our understanding of the fueling of star formation rate across cosmic time is incomplete without an understanding of how the contribution of \hi to the total composition of galaxies evolves with redshift, from the epoch of reionization until today. 

To infer the \hi\ gas content of even more distant galaxies, it is clear that a suitable tracer has to be identified as a proxy. This is similar to how observations of in particular CO and \ci has been used as tracers of the molecular gas content of galaxies through most of cosmic time \citep[e.g.][]{papadopoulos2004, bolatto2013}. We here investigate the $158\,{\rm \mu m}$ ($1900.5\,{\rm GHz}$) fine-structure transition of singly-ionized carbon ([C{\sc ii}]) as a potential tracer of \hi. This particular transition was suggested by \cite{morton1975} early on to be a strong cooling line within \hi\ regions. Further work by \cite{Tielens1985,Wolfire1995,Hollenbach1999, Wolfire2003} chemically examined the role of \cii in the cooling of photodissociation regions (PDRs) and the cold and warm neutral medium (CNM and WNM, respectively). Observationally, \cite{madden1993} found that extended \cii emission in NGC 6946 likely emerged from cold, atomic hydrogen clouds. \cite{madden1997} investigated a correlation between \cii luminosity and the \hi\ gas mass ($M_{\rm \hi}$) in the local irregular galaxy IC 10, finding that while most \cii emission arose from dense photodissociation regions (PDRs), $\sim 20\%$ of the emission could arise from atomic hydrogen regions in the interstellar medium (ISM) of the galaxy. More recent work relying on hydrodynamical simulations of galaxies \citep[e.g.,][]{vallini2015, lupi2020,RamosPadilla2021,RamosPadilla2022} have constrained the fraction of \cii originating from different ISM phases, showing that it is more predominantly produced in neutral \hi\ regions at higher redshifts. 
As \cii is one of the strongest cooling lines in the ISM, it is one of the brightest far-infrared lines and thus easily observable at high redshifts \citep{stacey2010}. This makes it a natural candidate as a potential proxy for \hi\ at $z\approx 6$. With the further evidence to support it as a tracer of \hi in the local universe, and the strong connection found in high-redshift gamma-ray burst (GRB) sightlines \citep{heintz2021}, we here wish to further explore the feasibility of \cii as a tracer of \hi\ based on simulations to a) clarify and further investigate the viability of \cii as a proxy for \hi, b) make predictions for the \hi\ gas mass fraction in galaxies at $z\approx 6$, c) examine the underlying physics which drive this correlation.

The paper is structured as follows:
In Section \ref{sec:sims+obs} we describe the observational and hydrodynamical simulation samples utilized for the purposes of this work. In Section \ref{sec:results} we describe the \cii{}-H{\scriptsize I} relation both in equation form and in terms of a single conversion factor, $\beta_{\rm \cii}$, and analyze the effect of metallicity on  $\beta_{\rm \cii}$. In Section \ref{sec:hib} we investigate the \hi gas budget at z $\approx$ 6. We conclude this work in Section \ref{sec:conclusion}.

\section{Simulations \& Observational Samples}\label{sec:sims+obs}

This work uses a set of \simba{} \citep{dave2019} galaxy simulations at $z = 5.93$, with line emission simulated and post-processed using v2 of \sigame{} \citep[][]{olsen2017}. The \simba simulation set consists of three cubical volumes of 25, 50 and $100\,{\rm cMpc}\,h^{-1}$ on a side, all of which are used in this work to search for galaxies at $z\sim6$. For each volume, a total of 1024$^3$ gas elements and 1024$^3$ dark matter particles are evolved from $z=249$. 
Within these three volumes, a total of 11,137 galaxies were selected via the yt \citep[][]{turk2011}-based package \caesar; \yt is a Python package used to visualize \& analyze volumetric data, and \caesar is a six-dimensional friends-of-friends algorithm which is applied to the simulated gas and stellar particles to identify and select \simba galaxies. The galaxy properties derived with \caesar include star formation rate (SFR), computed by dividing the stellar mass formed over a 100 Myr timescale \citep[][]{Leung2020}, SFR surface density, stellar mass ($M_{\rm \star}$), and a SFR-weighted gas-phase metallicity ($Z$). The ionized, molecular, and atomic gas masses are calculated via \caesar to be equal to the total mass of all gas particles associated with each respective phase \citep[we refer to][for further description of how these gas ``particles'' are defined and constructed]{olsen2017, vizgan2022}.

Our final sample consists of 11,125 galaxies, with \cii luminosities in the range $L_{\rm [CII]} = 10^{3.82}-10^{8.91}$\,\lsun{} and a molecular gas masses $M_{\rm mol} = 10^{6.25}-10^{10.33}$\,\msun{}, with a mean $M_{\rm \star}$ of 10$^{8.07}$ M$_{\rm \odot}$, mean SFR of 1.9 M$_{\rm \odot}/$yr, and mean metallicity of 0.18 Z$_{\rm \odot}$. We refer to \cite{Leung2020} and \cite{vizgan2022} for further description and detailing of \simba, \sigame, \caesar, and the simulation sample. 

In this work, we define the total \hi gas mass in each galaxy to be equal to the sum of its ionized and atomic hydrogen gas mass, independent of galactic region or radius. Molecular clouds have been shown observationally to also consist of a substantial fraction of \hi\ and will thereby also contribute to the total \hi gas mass. The fraction of \hi in molecular clouds has been measured to vary between 20 to 80\% in our Galaxy \citep[][]{burgh2010}, so here we assume that $\sim$50\% of the total molecular gas in our simulations is \hi. Similarly, a significant \hi volume filling fraction of $\approx 50\%$ has been derived for interstellar \hi within the warm neutral medium in the solar neighborhood \citep{heiles2003}. The way that \sigame classifies gas particles as ``ionized'' in this gas-phase is by requiring that its electron-to-hydrogen fraction is $x_{e^-} > 0.5$; consequently, a gas particle in \sigame will be considered completely ionized beyond $x_{e^-} > 0.5$, and otherwise neutral. This is not realistic when considering a real galaxy and we thus assume $\sim$50\% of the ionized gas will contribute to the atomic gas mass because of this classification. Because of this assumption, the gas masses derived from simulations should technically be treated as upper limits on the neutral gas budget.

We supplement our simulation sets with some observational samples of \cii{} observations from the literature. We use \hi{} gas masses \citep{cormier2015} of main-sequence dwarf galaxies with \cii{} observations in the local Herschel Dwarf Galaxy Survey \citep{madden2013}, which serve as high-redshift analogs. In addition, we use the X-shooter GRB afterglow legacy sample \citep[XS-GRB;][]{selsing2019}, which cover a redshift range between 0.059 and 7.84 and which were detected with the \textit{Swift} satellite. Metallicities and \hi{} column densities of these GRBs are taken from \cite{bolmer2019}, and the \cii{} column densities are taken from \cite{heintz2021}.

\section{The \cii-to-\hi relation}\label{sec:results}

\subsection{Connecting $L_{\rm \cii}$ and $M_{\rm HI}$}

\begin{figure}
    \centering
    \hspace*{-0.8cm}
    \includegraphics[width=10cm]{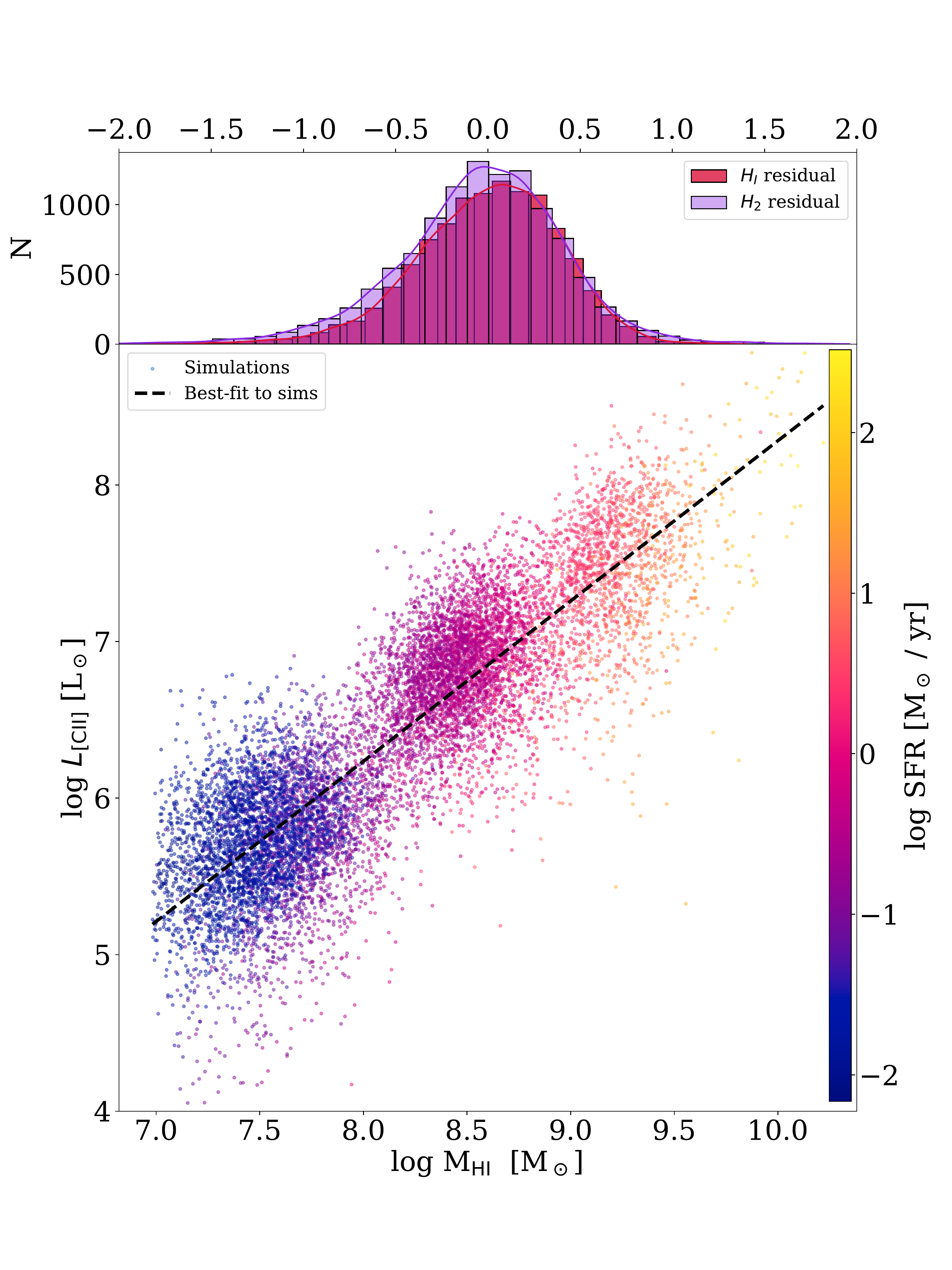}
    \caption{\textbf{Top panel:} Histograms of scatter around the best-fit models for the \cii-to-\hi relation (this work, crimson) and for the \cii-H$_2$ relation \citep[][violet]{vizgan2022}; the scatter around the former relation is tighter (0.39 dex) than for the \cii-H$_2$ relation (0.45 dex). \textbf{Bottom figure:} The \cii luminosity vs. \hi{} gas mass relation for our $z\simeq6$ simulations. We find a very strong linear correlation between these two quantities, yielding a best-fit of $\log L_{\rm{\cii}} = 1.02 \log M_{\rm{\hi}} - 1.95$.  The simulations are color-coded by the star formation rate, further demonstrating a correlation between SFR and the log-linear \cii-\hi{} relation. When accounting only for the diffuse \hi mass derived from \cloudy, the relation changes to $\log L_{\rm{\cii}} = 0.84 \log M_{\rm{\hi}} + 0.55$ with a scatter of 0.22 dex.}
    \label{fig:CII_HI}
\end{figure}

We examine the total \cii{} luminosity of the $z \simeq 6$ simulated galaxies and its connection to the total diffuse \hi gas mass in Fig.~\ref{fig:CII_HI}, color-coded as a function of SFR. 
In previous work, \cite{vizgan2022} attempted to test and verify the use of \cii as a tracer of the molecular gas based on the same set of simulated galaxies at $z\approx 6$. They found a sub-linear relation between $M_{\rm mol}$ and $L_{\rm \cii}$, in contrast with the linear relation described by \cite{zanella2018}, which arose as a consequence of the Kennicutt-Schmidt Law. By performing a log-linear fit to the data, we find a strong correlation between the \cii and \hi, yielding a best-fit relation of
\begin{equation}
   \log L_{\rm \cii}  = 1.02 \log M_{\rm \hi} - 1.95.
\end{equation}
with a root-mean-square (rms) scatter of 0.39 dex. For comparison, we examine the relation between $L_{\rm [CII]}$ and $M_{\rm mol}$ and SFR within these simulations. We find that scatter is notably tighter than $L_{\rm [CII]}$ vs. $M_{\rm mol}$ (0.45 dex) and than $L_{\rm [CII]}$ vs. SFR (0.50 dex) for our simulations \citep{vizgan2022}. This is further support of \cii being most tightly connected to \hi, and thus a more reliable tracer of this gas-phase. Figure 2 of \cite{vizgan2022} demonstrated that \cii emission was most negligible from \hi regions in these simulations; we highlight, though, that we want to calibrate the inferred HI mass with the total [CII] emission -- irrespective of origin --as this is what we ultimately observe from the high-z galaxies.

\subsection{The metallicity dependence of the \cii-to-\hi conversion factor}

\begin{figure}
    \centering
    \hspace*{-1.5cm}
    \includegraphics[width=10.5cm]{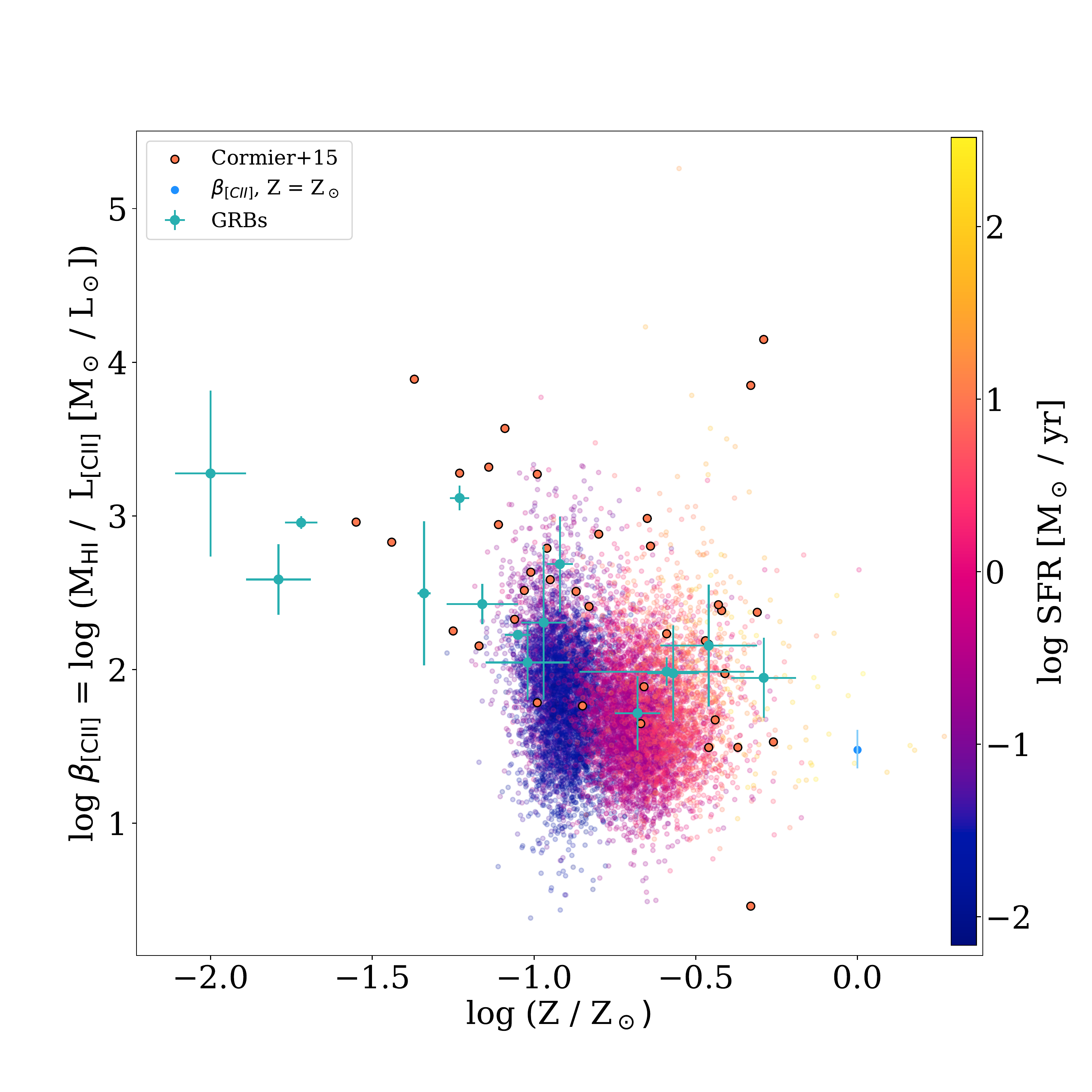}
    \caption{The [C{\scriptsize II}]-to-H{\scriptsize I} conversion factor $\beta_{[\rm CII]}$ vs. SFR-weighted metallicity. The blue dot is $\beta_{[\rm CII]} = 30^{+10}_{-7} \ \msun / \lsun$, the conversion factor at solar metallicities \citep{heintz2021}. GRBs are taken from \cite{selsing2019} \citep[see also][]{heintz2021}, while the orange points are the galaxies from \cite{cormier2015}.}
    \label{fig:beta_Zsfr}
\end{figure}

Defining the \cii-to-\hi conversion factor as $\beta_{\rm \cii} = M_{\rm \hi} / L_{\rm \cii}$ \citep{heintz2021}, we derive a median $\log \beta_{\rm \cii} = 1.7^{+0.4}_{-0.3}$ \msun / \lsun for our set of simulated galaxies at $z\approx 6$. \cite{heintz2021} determined that $\beta_{\rm \cii} = M_{\rm HI}/L_{\rm \cii}$ evolves with metallicity, following
\begin{multline}
    \log M_{\rm \hi} / L_{\rm \cii}  = (-0.87 \pm 0.09) \times \log \left(Z/Z_{\rm \odot}\right) 
    \\ + (1.48 \pm 0.12)
\label{eq:beta_vs_z_kasper}
\end{multline}
This \cii-to-\hi relation is derived from a combination of high-$z$ GRB afterglows, assuming that their pencil-beam sightlines probe representative regions of the star-forming ISM, and was found to show remarkable consistency with direct observations of local galaxies \citep{heintz2021}. 

In Fig.~\ref{fig:beta_Zsfr} we examine the evolution of $\beta_{\rm \cii}$ with metallicity within our simulations. For comparison, we overplot the measurements at $z=1.7-6.3$ from the GRB sightlines \citep{heintz2021} and the local set of observations \citep[][]{cormier2015}. We find an anti-correlation between $\beta_{\rm \cii}$ and $Z$ with our simulations (with r/pearson coefficient = $-$0.20), in strong agreement with \cite{heintz2021} and the local galaxy sample. Additionally, there appears to be a slight correlation between $\beta_{\rm \cii}$ and star formation rate. The anticorrelation of $\beta_{\rm \cii}$ is mostly a result of the definition; i.e. \cii/\hi is a proxy to carbon/hydrogen, hence the galaxies which are more-metal rich will have more carbon available, both in the neutral and ionized states.

\section{The \hi\ gas mass budget of galaxies at $z \approx 6$}\label{sec:hib}

With this set of simulations, we further make predictions of the \hi\ gas mass fraction of galaxies at $z\approx 6$. Defining the \hi\ gas mass excess as $M_{\rm \hi}/M_{\rm \star}$, we find a mean and median $M_{\rm \hi}/M_{\rm \star} = 1.4$ and 1.2$^{+0.9}_{-0.6}$, respectively. 
We also predict the fraction of \hi by mass to the total baryonic matter content $M_{\rm bar,tot} = M_{\rm \hi} + M_{\rm H_2} + M_{\rm \star}$. We find a mean and median fraction of $M_{\rm \hi} / M_{\rm bar,tot} = $ 0.45 and 0.46$^{+0.12}_{-0.13}$, respectively. By comparison, the mean $M_{\rm H_2}$/$M_{\rm bar,tot}$ is 0.17 and median 0.15$^{+0.10}_{-0.06}$; the mean $M_{\rm \star}$/$M_{\rm bar,tot}$ is 0.38, and the median is 0.37$^{+0.14}_{-0.11}$. We thus predict that \hi\ is the dominant, baryonic matter component in galaxies at $z\approx 6$.

Several observational works \citep[e.g.][]{zanella2018, dessauges2020} have investigated the use of \cii as a molecular gas mass tracer. \citet{vizgan2022} showed that the $\alpha_{\rm \cii}$ derived by \cite{zanella2018} (Z18) required further calibration, as the slope of the log-linear\cii-$M_{\rm mol}$ relation was not near unity. \cite{dessauges2020} used the Z18 $\alpha_{\rm \cii}$ relation to derive the molecular gas budgets of their observations across redshift. 

In Figure 3 we investigate the evolution of the \hi  gas mass budget, comparing the data from the ALPINE \cii survey \citep[e.g.][]{dessauges2020} with our simulations. The \hi gas masses in the ALPINE data are derived first by deriving metallicities from their SFRs and stellar masses via the high-redshift mass-metallicity relation in \cite{curti2020}, and then by using Equation 2 in this work \citep[also see][]{heintz2021}. We find that the ALPINE data are in reasonable agreement with the simulations; direct metallicity measurements from the ALMA sources, which were not available in the data, would allow for better constraints on the \hi gas mass. Nonetheless, the ALPINE data are consistent with the prediction from simulations that the \hi gas mass content should dominate the ISM at high-redshift. As the ALPINE team used the work from \cite{zanella2018} to derive molecular gas masses for their sources, further work at high-redshift should further examine the degree to which \cii luminosity traces the \hi, H$_{\rm 2}$, or both gas masses, as it appears to be a good tracer of both the neutral and molecular gas reservoirs which occupy the ISMs of high-redshift galaxies.

\begin{figure}
    \centering
    \hspace*{-1.5cm}
    \includegraphics[width=10cm]{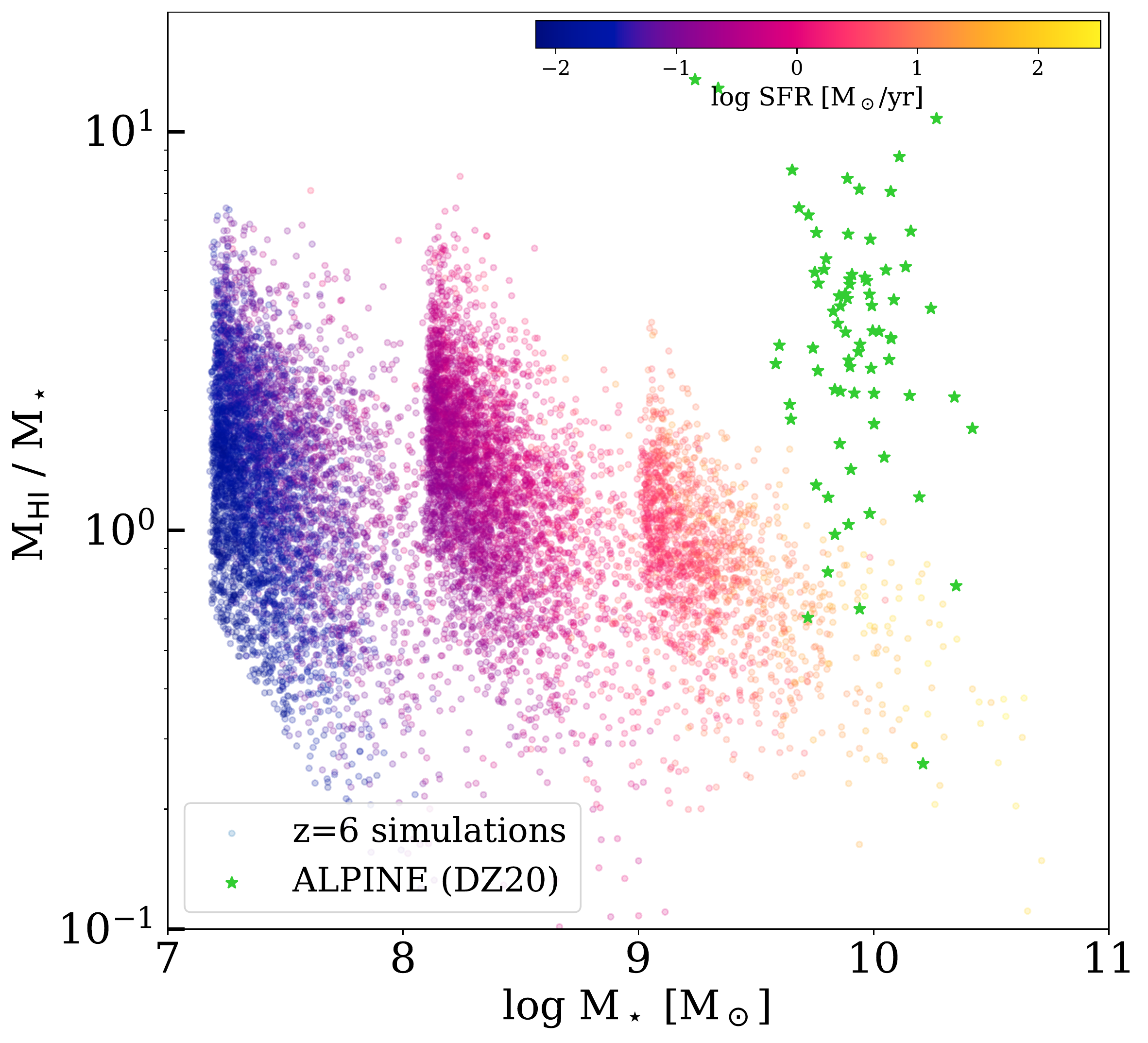}
    \caption{The evolution of the \hi gas mass excess with stellar mass. Our simulations are color-coded by star formation rate, and the y-axis is scaled logarithmically. The green stars represent galaxies from the ALPINE survey \cite{dessauges2020} For the ALPINE data, we derive gas-phase metallicities from their $M_{\rm \star}$ and SFRs using the mass-metallicity relation in \cite{curti2020}. From there, we use Equation 2 to derive estimates of their \hi gas mass content.}
    \label{fig:Mratio_logOH}
\end{figure}

\section{Conclusions}\label{sec:conclusion}

Previously, \cii has been used as a tracer of SFR in galaxies \citep[e.g.][]{delooze2014}, and more recently as a tracer of H$_2$ gas mass \citep[e.g.][]{zanella2018, madden2020, dessauges2020, vizgan2022}. In this work, we use a set of simulated galaxies at $z \simeq 6$ \citep[][]{vizgan2022} to investigate the utility of \cii luminosity as a tracer of atomic (i.e. diffuse) hydrogen gas mass, following the work of \cite{heintz2021}. We present our main findings below: 

\begin{enumerate}
    \item For our \simba simulations at $z=6$, we are able to find a strong relationship between \cii luminosity and \hi gas mass: $\log L_{\rm \cii} = 1.02 \log M_{\rm \hi} - 1.95$. Significantly, the relation is more tightly correlated (0.39 dex) than the \cii-to-$\rm{H_2}$ (0.45 dex) and \cii-to-SFR (0.50 dex) correlations described in \cite{vizgan2022}. Furthermore, we find an anti-correlation of of $\beta_{\rm \cii}$ with metallicity.
    
    \item We find that the \hi gas mass is a critical baryonic matter component, with mean $M_{\rm \hi}/M_{\rm \star} = 1.4$, and $M_{\rm \hi}/M_{\rm bar,tot} = 0.45$. We conclude that in the high-redshift universe, galaxies' ISMs (excluding dust) are dominated by \hi baryonic matter content. 
    
    \item We compare our work with archival \cii observational data, finding that extrapolating \hi gas masses from high-redshift \cii sources in the ALPINE \cii survey yields an evolution of the \hi gas mass budget over stellar mass which agrees with predictions from simulations.
    
\end{enumerate}

We ultimately conclude that \cii is a viable proxy to infer amount of \hi in galaxies at $z\approx 6$. \cii is readily available in the high-redshift universe for large surveys (e.g. the Large ALMA-ALPINE survey at $z\sim 4-6$ \citep[][etc.]{dessauges2020, lefevre2020, bethermin2020, faisst2020}; and REBELS at $z\sim 6-8$ \citep[e.g.;][]{bouwens2021, heintz2022}. Establishing a proxy for \hi\ is particularly important as even the next-generation radio facilities such as the SKA will only be able to detect and measure \hi out to z $\approx 2$, and will not be able to probe the evolution of \hi in the universe beyond cosmic noon. 

\section{Acknowledgements}

We are grateful the anonymous referee for insights and feedback which strengthened this work. We thank Robert Thompson for developing \caesar, and the \yt team
for development and support of \yt. 
This research made use of the the TOPCAT data analysis software \citep[][]{TOPCAT}.
This research was made possible by the National Science Foundation (NSF)-funded DAWN-IRES program 
and would not be possible without support from the Cosmic Dawn Center (DAWN) in Copenhagen, Denmark. 
The Cosmic Dawn Center (DAWN) is funded by the Danish National Research Foundation under grant No. 140.
DV was funded by an Open Study/Research Award from the Fulbright U.S. Student Program in Denmark.
KEH acknowledges support from the Carlsberg Foundation Reintegration Fellowship Grant CF21-0103.
KPO is funded by NASA under award No 80NSSC19K1651. 
DN acknowledges support from the NSF via grant AST 1909153. 
TRG acknowledges support from the Carlsberg Foundation
(grant no CF20-0534)
RD acknowledges support from the Wolfson Research Merit Award program of the U.K. Royal Society.
\simba was run on the DiRAC@Durham facility managed by the Institute for Computational Cosmology on behalf of the STFC DiRAC HPC Facility. The equipment was funded by BEIS capital funding via STFC capital grants ST/P002293/1, ST/R002371/1 and ST/S002502/1, Durham University and STFC operations grant ST/R000832/1. DiRAC is part of the National e-Infrastructure.

\bibliography{sample631}{}

\begin{thebibliography}{}
\expandafter\ifx\csname natexlab\endcsname\relax\def\natexlab#1{#1}\fi
\providecommand{\url}[1]{\href{#1}{#1}}
\providecommand{\dodoi}[1]{doi:~\href{http://doi.org/#1}{\nolinkurl{#1}}}
\providecommand{\doeprint}[1]{\href{http://ascl.net/#1}{\nolinkurl{http://ascl.net/#1}}}
\providecommand{\doarXiv}[1]{\href{https://arxiv.org/abs/#1}{\nolinkurl{https://arxiv.org/abs/#1}}}

\bibitem[{{Bera} {et~al.}(2019){Bera}, {Kanekar}, {Chengalur}, \&
  {Bagla}}]{bera2019}
{Bera}, A., {Kanekar}, N., {Chengalur}, J.~N., \& {Bagla}, J.~S. 2019, \apjl,
  882, L7, \dodoi{10.3847/2041-8213/ab3656}

\bibitem[{{B{\'e}thermin} {et~al.}(2020){B{\'e}thermin}, {Fudamoto}, {Ginolfi},
  {Loiacono}, {Khusanova}, {Capak}, {Cassata}, {Faisst}, {Le F{\`e}vre},
  {Schaerer}, {Silverman}, {Yan}, {Amorin}, {Bardelli}, {Boquien}, {Cimatti},
  {Davidzon}, {Dessauges-Zavadsky}, {Fujimoto}, {Gruppioni}, {Hathi}, {Ibar},
  {Jones}, {Koekemoer}, {Lagache}, {Lemaux}, {Moreau}, {Oesch}, {Pozzi},
  {Riechers}, {Talia}, {Toft}, {Vallini}, {Vergani}, {Zamorani}, \&
  {Zucca}}]{bethermin2020}
{B{\'e}thermin}, M., {Fudamoto}, Y., {Ginolfi}, M., {et~al.} 2020, \aap, 643,
  A2, \dodoi{10.1051/0004-6361/202037649}

\bibitem[{{Bolatto} {et~al.}(2013){Bolatto}, {Wolfire}, \&
  {Leroy}}]{bolatto2013}
{Bolatto}, A.~D., {Wolfire}, M., \& {Leroy}, A.~K. 2013, \araa, 51, 207,
  \dodoi{10.1146/annurev-astro-082812-140944}

\bibitem[{{Bolmer} {et~al.}(2019){Bolmer}, {Ledoux}, {Wiseman}, {De Cia},
  {Selsing}, {Schady}, {Greiner}, {Savaglio}, {Burgess}, {D'Elia}, {Fynbo},
  {Goldoni}, {Hartmann}, {Heintz}, {Jakobsson}, {Japelj}, {Kaper}, {Tanvir},
  {Vreeswijk}, \& {Zafar}}]{bolmer2019}
{Bolmer}, J., {Ledoux}, C., {Wiseman}, P., {et~al.} 2019, \aap, 623, A43,
  \dodoi{10.1051/0004-6361/201834422}

\bibitem[{{Bouwens} {et~al.}(2021){Bouwens}, {Smit}, {Schouws}, {Stefanon},
  {Bowler}, {Endsley}, {Gonzalez}, {Inami}, {Stark}, {Oesch}, {Hodge},
  {Aravena}, {da Cunha}, {Dayal}, {de Looze}, {Ferrara}, {Fudamoto},
  {Graziani}, {Li}, {Nanayakkara}, {Pallotini}, {Schneider}, {Sommovigo},
  {Topping}, {van der Werf}, {Algera}, {Barrufet}, {Hygate}, {Labbe},
  {Riechers}, \& {Witstok}}]{bouwens2021}
{Bouwens}, R.~J., {Smit}, R., {Schouws}, S., {et~al.} 2021, arXiv e-prints,
  arXiv:2106.13719.
\newblock \doarXiv{2106.13719}

\bibitem[{{Burgh} {et~al.}(2010){Burgh}, {France}, \& {Jenkins}}]{burgh2010}
{Burgh}, E.~B., {France}, K., \& {Jenkins}, E.~B. 2010, \apj, 708, 334,
  \dodoi{10.1088/0004-637X/708/1/334}

\bibitem[{{Carilli} \& {Walter}(2013)}]{carilli2013}
{Carilli}, C.~L., \& {Walter}, F. 2013, \araa, 51, 105,
  \dodoi{10.1146/annurev-astro-082812-140953}

\bibitem[{{Chowdhury} {et~al.}(2020){Chowdhury}, {Kanekar}, {Chengalur},
  {Sethi}, \& {Dwarakanath}}]{chowdhury2020}
{Chowdhury}, A., {Kanekar}, N., {Chengalur}, J.~N., {Sethi}, S., \&
  {Dwarakanath}, K.~S. 2020, \nat, 586, 369, \dodoi{10.1038/s41586-020-2794-7}

\bibitem[{{Chowdhury} {et~al.}(2021){Chowdhury}, {Kanekar}, {Das},
  {Dwarakanath}, \& {Sethi}}]{chowdhury2021}
{Chowdhury}, A., {Kanekar}, N., {Das}, B., {Dwarakanath}, K.~S., \& {Sethi}, S.
  2021, \apjl, 913, L24, \dodoi{10.3847/2041-8213/abfcc7}

\bibitem[{{Cormier} {et~al.}(2015){Cormier}, {Madden}, {Lebouteiller}, {Abel},
  {Hony}, {Galliano}, {R{\'e}my-Ruyer}, {Bigiel}, {Baes}, {Boselli},
  {Chevance}, {Cooray}, {De Looze}, {Doublier}, {Galametz}, {Hughes},
  {Karczewski}, {Lee}, {Lu}, \& {Spinoglio}}]{cormier2015}
{Cormier}, D., {Madden}, S.~C., {Lebouteiller}, V., {et~al.} 2015, \aap, 578,
  A53, \dodoi{10.1051/0004-6361/201425207}

\bibitem[{{Curti} {et~al.}(2020){Curti}, {Mannucci}, {Cresci}, \&
  {Maiolino}}]{curti2020}
{Curti}, M., {Mannucci}, F., {Cresci}, G., \& {Maiolino}, R. 2020, \mnras, 491,
  944, \dodoi{10.1093/mnras/stz2910}

\bibitem[{{Dav{\'e}} {et~al.}(2019){Dav{\'e}}, {Angl{\'e}s-Alc{\'a}zar},
  {Narayanan}, {Li}, {Rafieferantsoa}, \& {Appleby}}]{dave2019}
{Dav{\'e}}, R., {Angl{\'e}s-Alc{\'a}zar}, D., {Narayanan}, D., {et~al.} 2019,
  \mnras, 486, 2827, \dodoi{10.1093/mnras/stz937}

\bibitem[{{De Looze} {et~al.}(2014){De Looze}, {Cormier}, {Lebouteiller},
  {Madden}, {Baes}, {Bendo}, {Boquien}, {Boselli}, {Clements}, {Cortese},
  {Cooray}, {Galametz}, {Galliano}, {Graci{\'a}-Carpio}, {Isaak}, {Karczewski},
  {Parkin}, {Pellegrini}, {R{\'e}my-Ruyer}, {Spinoglio}, {Smith}, \&
  {Sturm}}]{delooze2014}
{De Looze}, I., {Cormier}, D., {Lebouteiller}, V., {et~al.} 2014, \aap, 568,
  A62

\bibitem[{{Dessauges-Zavadsky} {et~al.}(2020){Dessauges-Zavadsky}, {Ginolfi},
  {Pozzi}, {B{\'e}thermin}, {Le F{\`e}vre}, {Fujimoto}, {Silverman}, {Jones},
  {Vallini}, {Schaerer}, {Faisst}, {Khusanova}, {Fudamoto}, {Cassata},
  {Loiacono}, {Capak}, {Yan}, {Amorin}, {Bardelli}, {Boquien}, {Cimatti},
  {Gruppioni}, {Hathi}, {Ibar}, {Koekemoer}, {Lemaux}, {Narayanan}, {Oesch},
  {Rodighiero}, {Romano}, {Talia}, {Toft}, {Vergani}, {Zamorani}, \&
  {Zucca}}]{dessauges2020}
{Dessauges-Zavadsky}, M., {Ginolfi}, M., {Pozzi}, F., {et~al.} 2020, \aap, 643,
  A5

\bibitem[{{Faisst} {et~al.}(2020){Faisst}, {Schaerer}, {Lemaux}, {Oesch},
  {Fudamoto}, {Cassata}, {B{\'e}thermin}, {Capak}, {Le F{\`e}vre}, {Silverman},
  {Yan}, {Ginolfi}, {Koekemoer}, {Morselli}, {Amor{\'\i}n}, {Bardelli},
  {Boquien}, {Brammer}, {Cimatti}, {Dessauges-Zavadsky}, {Fujimoto},
  {Gruppioni}, {Hathi}, {Hemmati}, {Ibar}, {Jones}, {Khusanova}, {Loiacono},
  {Pozzi}, {Talia}, {Tasca}, {Riechers}, {Rodighiero}, {Romano}, {Scoville},
  {Toft}, {Vallini}, {Vergani}, {Zamorani}, \& {Zucca}}]{faisst2020}
{Faisst}, A.~L., {Schaerer}, D., {Lemaux}, B.~C., {et~al.} 2020, \apjs, 247,
  61, \dodoi{10.3847/1538-4365/ab7ccd}

\bibitem[{{Fern{\'a}ndez} {et~al.}(2016){Fern{\'a}ndez}, {Gim}, {van Gorkom},
  {Yun}, {Momjian}, {Popping}, {Chomiuk}, {Hess}, {Hunt}, {Kreckel}, {Lucero},
  {Maddox}, {Oosterloo}, {Pisano}, {Verheijen}, {Hales}, {Chung}, {Dodson},
  {Golap}, {Gross}, {Henning}, {Hibbard}, {Jaff{\'e}}, {Donovan Meyer},
  {Meyer}, {Sanchez-Barrantes}, {Schiminovich}, {Wicenec}, {Wilcots},
  {Bershady}, {Scoville}, {Strader}, {Tremou}, {Salinas}, \&
  {Ch{\'a}vez}}]{fernandez2016}
{Fern{\'a}ndez}, X., {Gim}, H.~B., {van Gorkom}, J.~H., {et~al.} 2016, \apjl,
  824, L1, \dodoi{10.3847/2041-8205/824/1/L1}

\bibitem[{{Heiles} \& {Troland}(2003)}]{heiles2003}
{Heiles}, C., \& {Troland}, T.~H. 2003, \apj, 586, 1067, \dodoi{10.1086/367828}

\bibitem[{{Heintz} {et~al.}(2021){Heintz}, {Watson}, {Oesch}, {Narayanan}, \&
  {Madden}}]{heintz2021}
{Heintz}, K.~E., {Watson}, D., {Oesch}, P., {Narayanan}, D., \& {Madden}, S.~C.
  2021, arXiv e-prints, arXiv:2108.13442.
\newblock \doarXiv{2108.13442}

\bibitem[{{Heintz} {et~al.}(2022){Heintz}, {Oesch}, {Aravena}, {Bouwens},
  {Dayal}, {Ferrara}, {Fudamoto}, {Graziani}, {Inami}, {Sommovigo}, {Smit},
  {Stefanon}, {Topping}, {Pallottini}, \& {van der Werf}}]{heintz2022}
{Heintz}, K.~E., {Oesch}, P.~A., {Aravena}, M., {et~al.} 2022, arXiv e-prints,
  arXiv:2206.07763.
\newblock \doarXiv{2206.07763}

\bibitem[{{Hollenbach} \& {Tielens}(1999)}]{Hollenbach1999}
{Hollenbach}, D.~J., \& {Tielens}, A.~G.~G.~M. 1999, Reviews of Modern Physics,
  71, 173, \dodoi{10.1103/RevModPhys.71.173}

\bibitem[{{Kalberla} {et~al.}(2005){Kalberla}, {Burton}, {Hartmann}, {Arnal},
  {Bajaja}, {Morras}, \& {P{\"o}ppel}}]{kalberla2005}
{Kalberla}, P.~M.~W., {Burton}, W.~B., {Hartmann}, D., {et~al.} 2005, \aap,
  440, 775, \dodoi{10.1051/0004-6361:20041864}

\bibitem[{{Le F{\`e}vre} {et~al.}(2020){Le F{\`e}vre}, {B{\'e}thermin},
  {Faisst}, {Jones}, {Capak}, {Cassata}, {Silverman}, {Schaerer}, {Yan},
  {Amorin}, {Bardelli}, {Boquien}, {Cimatti}, {Dessauges-Zavadsky},
  {Giavalisco}, {Hathi}, {Fudamoto}, {Fujimoto}, {Ginolfi}, {Gruppioni},
  {Hemmati}, {Ibar}, {Koekemoer}, {Khusanova}, {Lagache}, {Lemaux}, {Loiacono},
  {Maiolino}, {Mancini}, {Narayanan}, {Morselli}, {M{\'e}ndez-Hern{\`a}ndez},
  {Oesch}, {Pozzi}, {Romano}, {Riechers}, {Scoville}, {Talia}, {Tasca},
  {Thomas}, {Toft}, {Vallini}, {Vergani}, {Walter}, {Zamorani}, \&
  {Zucca}}]{lefevre2020}
{Le F{\`e}vre}, O., {B{\'e}thermin}, M., {Faisst}, A., {et~al.} 2020, \aap,
  643, A1

\bibitem[{{Leung} {et~al.}(2020){Leung}, {Olsen}, {Somerville}, {Dav{\'e}},
  {Greve}, {Hayward}, {Narayanan}, \& {Popping}}]{Leung2020}
{Leung}, T.~K.~D., {Olsen}, K.~P., {Somerville}, R.~S., {et~al.} 2020, \apj,
  905, 102

\bibitem[{{Lupi} \& {Bovino}(2020)}]{lupi2020}
{Lupi}, A., \& {Bovino}, S. 2020, \mnras, 492, 2818,
  \dodoi{10.1093/mnras/staa048}

\bibitem[{{Madau} \& {Dickinson}(2014)}]{madau2014}
{Madau}, P., \& {Dickinson}, M. 2014, \araa, 52, 415,
  \dodoi{10.1146/annurev-astro-081811-125615}

\bibitem[{{Madden} {et~al.}(1993){Madden}, {Geis}, {Genzel}, {Herrmann},
  {Jackson}, {Poglitsch}, {Stacey}, \& {Townes}}]{madden1993}
{Madden}, S.~C., {Geis}, N., {Genzel}, R., {et~al.} 1993, \apj, 407, 579,
  \dodoi{10.1086/172539}

\bibitem[{{Madden} {et~al.}(1997){Madden}, {Poglitsch}, {Geis}, {Stacey}, \&
  {Townes}}]{madden1997}
{Madden}, S.~C., {Poglitsch}, A., {Geis}, N., {Stacey}, G.~J., \& {Townes},
  C.~H. 1997, \apj, 483, 200, \dodoi{10.1086/304247}

\bibitem[{{Madden} {et~al.}(2013){Madden}, {R{\'e}my-Ruyer}, {Galametz},
  {Cormier}, {Lebouteiller}, {Galliano}, {Hony}, {Bendo}, {Smith}, {Pohlen},
  {Roussel}, {Sauvage}, {Wu}, {Sturm}, {Poglitsch}, {Contursi}, {Doublier},
  {Baes}, {Barlow}, {Boselli}, {Boquien}, {Carlson}, {Ciesla}, {Cooray},
  {Cortese}, {de Looze}, {Irwin}, {Isaak}, {Kamenetzky}, {Karczewski}, {Lu},
  {MacHattie}, {O'Halloran}, {Parkin}, {Rangwala}, {Schirm}, {Schulz},
  {Spinoglio}, {Vaccari}, {Wilson}, \& {Wozniak}}]{madden2013}
{Madden}, S.~C., {R{\'e}my-Ruyer}, A., {Galametz}, M., {et~al.} 2013, \pasp,
  125, 600, \dodoi{10.1086/671138}

\bibitem[{{Madden} {et~al.}(2020){Madden}, {Cormier}, {Hony}, {Lebouteiller},
  {Abel}, {Galametz}, {De Looze}, {Chevance}, {Polles}, {Lee}, {Galliano},
  {Lambert-Huyghe}, {Hu}, \& {Ramambason}}]{madden2020}
{Madden}, S.~C., {Cormier}, D., {Hony}, S., {et~al.} 2020, \aap, 643, A141,
  \dodoi{10.1051/0004-6361/202038860}

\bibitem[{{Morton} \& {Hu}(1975)}]{morton1975}
{Morton}, D.~C., \& {Hu}, E.~M. 1975, \apj, 202, 638, \dodoi{10.1086/154019}

\bibitem[{{Olsen} {et~al.}(2017){Olsen}, {Greve}, {Narayanan}, {Thompson},
  {Dav{\'e}}, {Niebla Rios}, \& {Stawinski}}]{olsen2017}
{Olsen}, K., {Greve}, T.~R., {Narayanan}, D., {et~al.} 2017, \apj, 846, 105

\bibitem[{{Papadopoulos} \& {Greve}(2004)}]{papadopoulos2004}
{Papadopoulos}, P.~P., \& {Greve}, T.~R. 2004, \apjl, 615, L29,
  \dodoi{10.1086/426059}

\bibitem[{{Ramos Padilla} {et~al.}(2021){Ramos Padilla}, {Wang}, {Ploeckinger},
  {van der Tak}, \& {Trager}}]{RamosPadilla2021}
{Ramos Padilla}, A.~F., {Wang}, L., {Ploeckinger}, S., {van der Tak}, F.~F.~S.,
  \& {Trager}, S.~C. 2021, \aap, 645, A133, \dodoi{10.1051/0004-6361/202038207}

\bibitem[{{Ramos Padilla} {et~al.}(2022){Ramos Padilla}, {Wang}, {van der Tak},
  \& {Trager}}]{RamosPadilla2022}
{Ramos Padilla}, A.~F., {Wang}, L., {van der Tak}, F.~F.~S., \& {Trager}, S.
  2022, arXiv e-prints, arXiv:2205.11955.
\newblock \doarXiv{2205.11955}

\bibitem[{{Scott} \& {Rees}(1990)}]{scott1990}
{Scott}, D., \& {Rees}, M.~J. 1990, \mnras, 247, 510

\bibitem[{{Selsing} {et~al.}(2019){Selsing}, {Malesani}, {Goldoni}, {Fynbo},
  {Kr{\"u}hler}, {Antonelli}, {Arabsalmani}, {Bolmer}, {Cano}, {Christensen},
  {Covino}, {D'Avanzo}, {D'Elia}, {De Cia}, {de Ugarte Postigo}, {Flores},
  {Friis}, {Gomboc}, {Greiner}, {Groot}, {Hammer}, {Hartoog}, {Heintz},
  {Hjorth}, {Jakobsson}, {Japelj}, {Kann}, {Kaper}, {Ledoux}, {Leloudas},
  {Levan}, {Maiorano}, {Melandri}, {Milvang-Jensen}, {Palazzi}, {Palmerio},
  {Perley}, {Pian}, {Piranomonte}, {Pugliese}, {S{\'a}nchez-Ram{\'\i}rez},
  {Savaglio}, {Schady}, {Schulze}, {Sollerman}, {Sparre}, {Tagliaferri},
  {Tanvir}, {Th{\"o}ne}, {Vergani}, {Vreeswijk}, {Watson}, {Wiersema},
  {Wijers}, {Xu}, \& {Zafar}}]{selsing2019}
{Selsing}, J., {Malesani}, D., {Goldoni}, P., {et~al.} 2019, \aap, 623, A92,
  \dodoi{10.1051/0004-6361/201832835}

\bibitem[{{Stacey} {et~al.}(2010){Stacey}, {Hailey-Dunsheath}, {Ferkinhoff},
  {Nikola}, {Parshley}, {Benford}, {Staguhn}, \& {Fiolet}}]{stacey2010}
{Stacey}, G.~J., {Hailey-Dunsheath}, S., {Ferkinhoff}, C., {et~al.} 2010, \apj,
  724, 957, \dodoi{10.1088/0004-637X/724/2/957}

\bibitem[{{Swaters} {et~al.}(2002){Swaters}, {van Albada}, {van der Hulst}, \&
  {Sancisi}}]{swaters2002}
{Swaters}, R.~A., {van Albada}, T.~S., {van der Hulst}, J.~M., \& {Sancisi}, R.
  2002, \aap, 390, 829, \dodoi{10.1051/0004-6361:20011755}

\bibitem[{{Taylor}(2005)}]{TOPCAT}
{Taylor}, M.~B. 2005, in Astronomical Society of the Pacific Conference Series,
  Vol. 347, Astronomical Data Analysis Software and Systems XIV, ed.
  P.~{Shopbell}, M.~{Britton}, \& R.~{Ebert}, 29

\bibitem[{{Thuan} \& {Martin}(1981)}]{thuan1981}
{Thuan}, T.~X., \& {Martin}, G.~E. 1981, \apj, 247, 823, \dodoi{10.1086/159094}

\bibitem[{{Tielens} \& {Hollenbach}(1985)}]{Tielens1985}
{Tielens}, A.~G.~G.~M., \& {Hollenbach}, D. 1985, \apj, 291, 722,
  \dodoi{10.1086/163111}

\bibitem[{{Turk} {et~al.}(2011){Turk}, {Smith}, {Oishi}, {Skory}, {Skillman},
  {Abel}, \& {Norman}}]{turk2011}
{Turk}, M.~J., {Smith}, B.~D., {Oishi}, J.~S., {et~al.} 2011, \apjs, 192, 9

\bibitem[{{Vallini} {et~al.}(2015){Vallini}, {Gallerani}, {Ferrara},
  {Pallottini}, \& {Yue}}]{vallini2015}
{Vallini}, L., {Gallerani}, S., {Ferrara}, A., {Pallottini}, A., \& {Yue}, B.
  2015, \apj, 813, 36, \dodoi{10.1088/0004-637X/813/1/36}

\bibitem[{{Vizgan} {et~al.}(2022){Vizgan}, {Greve}, {Olsen}, {Zanella},
  {Narayanan}, {Dav{\`e}}, {Magdis}, {Popping}, {Valentino}, \&
  {Heintz}}]{vizgan2022}
{Vizgan}, D., {Greve}, T.~R., {Olsen}, K.~P., {et~al.} 2022, \apj, 929, 92,
  \dodoi{10.3847/1538-4357/ac5cba}

\bibitem[{{Walter} {et~al.}(2008){Walter}, {Brinks}, {de Blok}, {Bigiel},
  {Kennicutt}, {Thornley}, \& {Leroy}}]{walter2008}
{Walter}, F., {Brinks}, E., {de Blok}, W.~J.~G., {et~al.} 2008, \aj, 136, 2563,
  \dodoi{10.1088/0004-6256/136/6/2563}

\bibitem[{{Wolfire} {et~al.}(1995){Wolfire}, {Hollenbach}, {McKee}, {Tielens},
  \& {Bakes}}]{Wolfire1995}
{Wolfire}, M.~G., {Hollenbach}, D., {McKee}, C.~F., {Tielens}, A.~G.~G.~M., \&
  {Bakes}, E.~L.~O. 1995, \apj, 443, 152, \dodoi{10.1086/175510}

\bibitem[{{Wolfire} {et~al.}(2003){Wolfire}, {McKee}, {Hollenbach}, \&
  {Tielens}}]{Wolfire2003}
{Wolfire}, M.~G., {McKee}, C.~F., {Hollenbach}, D., \& {Tielens}, A.~G.~G.~M.
  2003, \apj, 587, 278, \dodoi{10.1086/368016}

\bibitem[{{Zanella} {et~al.}(2018){Zanella}, {Daddi}, {Magdis}, {Diaz Santos},
  {Cormier}, {Liu}, {Cibinel}, {Gobat}, {Dickinson}, {Sargent}, {Popping},
  {Madden}, {Bethermin}, {Hughes}, {Valentino}, {Rujopakarn}, {Pannella},
  {Bournaud}, {Walter}, {Wang}, {Elbaz}, \& {Coogan}}]{zanella2018}
{Zanella}, A., {Daddi}, E., {Magdis}, G., {et~al.} 2018, \mnras, 481, 1976

\end{thebibliography}
\bibliographystyle{aasjournal}

\end{document}